\documentclass[aps,prc,superscriptaddress,showpacs,preprint]{revtex4}
\usepackage{graphicx}

\begin{document}

\title{Pairing-excitation versus intruder states in $^{68}$Ni and $^{90}$Zr}

 \author{D. Pauwels}
 \email[]{Dieter.Pauwels@fys.kuleuven.be}
 \affiliation{Instituut voor Kern- en Stralingsfysica, K.U. Leuven,
 Celestijnenlaan 200D, B-3001 Leuven, Belgium}
 \author{J.L. Wood}
 \affiliation{School of Physics, Georgia Institute of Technology, Atlanta, Georgia 30332-0430, USA}
 \author{K. Heyde}
 \affiliation{Department of Physics and Astronomy, Proeftuinstraat 86, B-9000 Gent, Belgium}
 \author{M. Huyse}
 \affiliation{Instituut voor Kern- en Stralingsfysica, K.U. Leuven,
 Celestijnenlaan 200D, B-3001 Leuven, Belgium}
 \author{R. Julin}
 \affiliation{University of Jyv\"{a}skyl\"{a}, Department of Physics, PO Box 35, FI-40014 Jyv\"{a}skyl\"{a}, Finland}
 \author{P. Van Duppen}
 \affiliation{Instituut voor Kern- en Stralingsfysica, K.U. Leuven,
 Celestijnenlaan 200D, B-3001 Leuven, Belgium}

\date{\today}

\begin{abstract}
A discussion on the nature of the $0^{+}$ states in $^{68}$Ni ($Z=28$, $N=40$) is presented and a comparison is made with its valence counterpart $^{90}$Zr ($Z=40$, $N=50$). Evidence is given for a $0^{+}$ proton intruder state at only $\sim 2.2$ MeV excitation energy in $^{68}$Ni, while the analogous neutron intruder states in $^{90}$Zr reside at $4126$ keV and $5441$ keV. The application of a shell-model description of $0^{+}$ intruder states reveals that many pair-scattered neutrons across $N=40$ have to be involved to explain the low excitation energy of the proton-intruder configuration in $^{68}$Ni.
\end{abstract}

\pacs{21.10.-k, 21.60.Cs, 27.50.+e, 27.60.+j}

\maketitle

\section{Introduction}

The nucleus $^{68}$Ni was initially considered as a semi-magic nucleus arising from a major $Z=28$ proton-shell closure and a $N=40$ neutron subshell closure. This interpretation was inferred from the high energy of the first-excited $2^{+}$ state ($2033$ keV \cite{Bro_PRL_95}) in contrast with the low energy of the first-excited $0^{+}$ state ($1770$ keV \cite{Ber_PL_82}). Conflicting observations arose, however, as mass measurements do not reveal a clear neutron shell gap at $N=40$ \cite{Rah_EPJ_07,Gue_PRC_07} and the $B(E2;0^{+}_{1} \rightarrow 2^{+}_{1})$ mean value of $3.2(6)$ W.u.~\cite{Sor_PRL_02,Bre_PRC_08} is too large for a pronounced $N=40$ subshell gap \cite{Lan_PRC_03}.

Currently, it is qualitatively understood that the apparent semi-magic properties of $^{68}$Ni are not caused by a strong $N=40$ subshell closure and a corresponding large energy gap, but rather follow from the parity change between the $pf$ shell and the $1g_{9/2}$ orbital across $N=40$, prohibiting quadrupole excitations \cite{Gra_AIP_00}. The $B(E2)$ value is explained by strong pair scattering across $N=40$ \cite{Sor_PRL_02}, which indicates that the stabilizing effect is subtle.

Despite these qualitative insights, the structure of $^{68}$Ni and the region around is not yet fully understood. While the focus was, so far, mainly on neutron excitations across $N=40$, little is known about proton excitations across $Z=28$. Although separation energies give evidence for a major $Z=28$ shell closure at $N=40$, a proton two-particle-two-hole $\pi$(2p-2h) $0^{+}$ state could appear nonetheless at lower excitation energies due to pairing correlations and proton-neutron $\pi$-$\nu$ residual interactions \cite{Hey_PR_83,Hey_NPA_87,Woo_PR_92}. Its excitation energy will depend critically, however, on the stabilizing properties of the $N=40$ gap as the quadrupole part of the $\pi$-$\nu$ interaction depends on the number of valence neutron particles or holes.

\begin{figure*}
\centering
\includegraphics[width=0.75\linewidth]{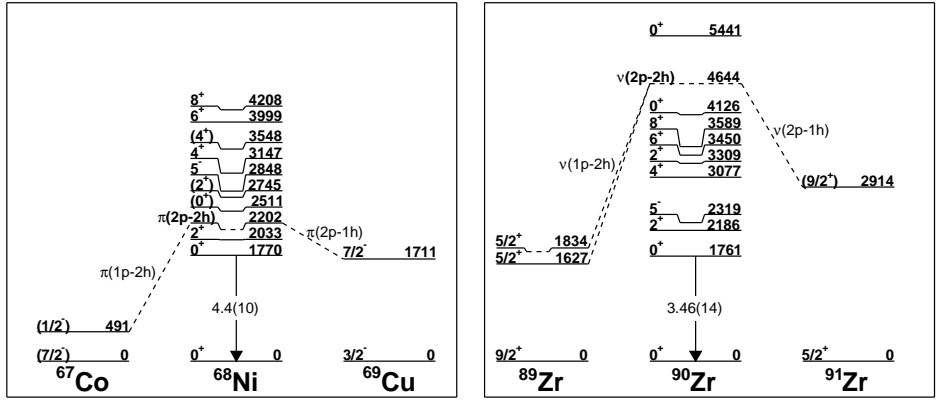}
\caption{Low-energy structure of $^{68}$Ni and $^{90}$Zr \cite{NNDC,Ber_PL_82,Mue_PRC_00,Bal_PRC_71}. The arrows denote $\rho^{2}$(E$0$)$\cdot 10^{3}$ transition strengths \cite{Kib_ADN_05}. The estimated excitation energies of the respective $\pi$ and $\nu$(2p-2h) configurations, based on 1p-2h and 2p-1h excitation energies of the $Z\pm1$ and $N\pm1$ nuclei \cite{Pau_PRC_08,Zei_PRC_78,Oro_NPA_00,Ste_PRL_08,Bal_PRC_72,Gle_NPA_71,Blo_NPA_76}, respectively, are represented by the dashed lines.} \label{fig:68Ni_90Zr_zeroplus}
\end{figure*}

Since the valence counterpart of $^{68}$Ni, $^{90}$Zr ($Z=40$, $N=50$), is a stable isotope, it has been investigated in numerous transfer reactions and thus its structure is better known than the one of $^{68}$Ni. In the present paper, the low-energy structures of both nuclei, and the $0^{+}$ states in particular, are compared based on experimental information available in the literature (see Fig.~\ref{fig:68Ni_90Zr_zeroplus}). While most properties are similar in $^{68}$Ni and  $^{90}$Zr, possible $\pi$(2p-2h) excitations in $^{68}$Ni will behave different from the $\nu$(2p-2h) excitations in $^{90}$Zr. In the following, a candidate for a $\pi$(2p-2h) $0^{+}$ is discussed on the basis of a shell-model approach of intruder states \cite{Hey_NPA_87}, after which implications for the stabilizing properties of the $N,Z=40$ gaps are discussed.

\section{Low-energy structure of $^{68}$Ni and $^{90}$Zr}\label{sec:zeroplus}

\subsection{The active valence nucleons}

At low excitation energies, the $^{68}$Ni and $^{90}$Zr valence nucleons (neutrons and protons, respectively) are expected to be predominantly active in the $2p_{1/2}$ and $1g_{9/2}$ space, and to a smaller degree in the $1f_{5/2}$ and $2p_{3/2}$ space. The energy difference between the $2p_{1/2}$ and $1g_{9/2}$ orbitals constitutes the $N,Z=40$ energy gap in $^{68}$Ni and $^{90}$Zr, respectively.

In a simplified picture, the ground state of $^{68}$Ni and $^{90}$Zr is expected to exhibit a $(2p_{3/2})^{4} (1f_{5/2})^{6} (2p_{1/2})^{2}$ character, while excited $0^{+}$ states could be created by promoting a nucleon pair from the $2p_{1/2}$ or $1f_{5/2}$ to the $1g_{9/2}$ orbital. It has been observed that the $0^{+}_{2}$ state in $^{68}$Ni and $^{90}$Zr feature remarkable similarities. Their respective excitation energies of $1770$ and $1761$ keV are almost identical, as well as their respective monopole $\rho^{2}$(E$0$) transition strengths of $4.4(10)\cdot 10^{-3}$ and $3.46(14)\cdot 10^{-3}$ \cite{Kib_ADN_05}.

Using spectroscopic factors from transfer reactions \cite{Cat_PR_69}, and mean-square radii $\langle r^{2} \rangle_{2p_{1/2}}$ and $\langle r^{2} \rangle_{1g_{9/2}}$ determined from the $^{90}$Zr($t$,$\alpha$)$^{89}$Y reaction \cite{War_PLB_79}, the measured $\rho^{2}$(E$0$) transition strength in $^{90}$Zr can be reproduced with a simple two-component model allowing for strong $\pi 2p_{1/2}$ ($59 \%$) and $\pi 1g_{9/2}$ ($41 \%$) configuration mixing \cite{Woo_NPA_99}. This gives substantial evidence for a strongly mixed $0_{1}^{+}$ ground state and $0_{2}^{+}$ excited state. Recent shell-model calculations confirm these observations \cite{Sie_PRC_09,Sie_PRC_09E}. Although the similar $\rho^{2}$(E$0$) transition strength in $^{68}$Ni is not understood, the similar excitation energy suggests comparable $0^{+}_{1}$ and $0^{+}_{2}$ configurations, involving now the neutrons.

The $0^{+}$ state arising from $(1f_{5/2})^{-2}$ has not been identified in $^{68}$Ni nor in $^{90}$Zr.  The $(0^{+}_{3})$ state, which has been observed in $^{68}$Ni at $2511$ keV \cite{Mue_PRC_00}, might be a possible candidate, although such a state is not observed in $^{90}$Zr in spite of the more extensive spectroscopic information.

The $2^{+}$, $4^{+}$, $6^{+}$, and $8^{+}$ levels in $^{68}$Ni at respective excitation energies of $2033$, $3147$, $3999$, and $4208$~keV are good candidates for the $g_{9/2}$ $v=2$ seniority levels. In $^{90}$Zr, a similar structure is observed with the respective $v=2$ seniority levels at excitation energies of $2186$, $3077$, $3450$, and $3589$ keV.

\subsection{Intruders across the $Z=28$ or $N=50$ gap}

The excitation energy of 2p-2h $0^{+}$ intruder states in nuclei at a major closed shell $Z$ (or $N$) can be estimated from summing the $\pi$($\nu$)(2p-1h) and $\pi$($\nu$)(1p-2h) intruder excitation energies in the $Z+1$($N+1$) and $Z-1$($N-1$) nuclei \cite{Van_PRL_84} (see Ref.~\cite{Hey_NPA_88} for details). Using this prescription, the excitation energies of, e.g., $\pi$(2p-2h) $0^{+}$ states in $Z=82$ lead and $Z=50$ tin nuclei are generally reproduced within $100$ keV.

In $^{89}$Zr, it is shown by the $^{91}$Zr(p,t) reaction \cite{Bal_PRC_72} that the $\nu$(1p-2h) configuration is mainly distributed over two states at excitation energies of $1627$ and $1834$~keV \cite{NNDC}. The $^{88}$Sr($\alpha$,n$\gamma$) \cite{Gle_NPA_71}, ($p$,$p^{'}$), and $^{92}$Zr($p$,$d$) reactions \cite{Blo_NPA_76} show that the major fraction of the $\nu$(2p-1h) configuration in $^{91}$Zr resides in the $2914$-keV state \cite{NNDC}. By using the above mentioned prescription and averaging the excitation energies of the two $\nu$(1p-2h) $^{89}$Zr levels, an expected excitation energy of $4644$~keV for the $\nu$(2p-2h) state in $^{90}$Zr can be deduced. The situation is depicted by the dashed lines in Fig.~\ref{fig:68Ni_90Zr_zeroplus}.

It has been shown by a $^{92}$Zr($p$,$t$) reaction \cite{Bal_PRC_71} that the $\nu$(2p-2h) configuration is mainly concentrated in the $0^{+}$ states at $4126$ and $5441$ keV excitation energy. The average of both excitation energies is $4784$~keV, which differs only by $140$~keV from the estimate.

The same reasoning can be applied to $^{68}$Ni. The $\pi$(2p-1h) character of the $1711$-keV level in $^{69}$Cu is suggested by a large spectroscopic factor in the $^{70}$Zn($d$,$^{3}$He) reaction \cite{Zei_PRC_78} and a small B(E2) transition strength to the $^{69}$Cu ground state observed in Coulomb excitation \cite{Ste_PRL_08}. From a recent $^{67}$Fe $\beta$-decay study \cite{Pau_PRC_08}, the $\pi$(1p-2h) state in $^{67}$Co was identified at $491$ keV, giving rise to an estimated excitation energy of the $\pi$(2p-2h) $0^{+}$ state in $^{68}$Ni at only $2202$ keV.

A good candidate for a $\pi$(2p-2h) $0^{+}$ configuration would be the $(0^{+}_{3})$ state in $^{68}$Ni, which is also a possible candidate for a $\nu 1f_{5/2}^{-2}$ state. From the presently available experimental data, however, it is not possible to differentiate between the two possible configurations. Although extremely challenging, future transfer and multi-Coulomb-excitation experiments can deliver crucial information to investigate this state and other low-energy levels in $^{68}$Ni.

In spite of their very similar excitation spectrum, there is thus a large difference in excitation energy of the 2p-2h intruder states across $Z=28$ or $N=50$ in respectively $^{68}$Ni and $^{90}$Zr. The possible reasons for this difference will now be investigated.

\section{Shell-model description for $\text{2p-2h}$ $0^{+}$ states in $^{68}$Ni and $^{90}$Zr}\label{sec:origin}

The $0^{+}_{2}$ excitation energies in $^{68}$Ni and $^{90}$Zr suggest nearly identic structures of their $0_{1}^{+}$ and $0_{2}^{+}$ states. On the other hand, the summed excitation energy of the $\pi$(1p-2h) and $\pi$(2p-1h) levels in $^{67}$Co and $^{69}$Cu, respectively, is very different from the $\nu$(2p-2h) excitation energy in $^{90}$Zr. The shell-model approach of Ref.~\cite{Hey_NPA_87} provides a quantitative description of $\pi$(2p-2h) and $\nu$(2p-2h) $0^{+}$ states, which can explain this apparent paradox in $^{68}$Ni and $^{90}$Zr.

\subsection{Framework and results}

Intruder states result from particle-hole excitations across major closed shells. Nevertheless, they appear at low excitation energy because of both strong pairing and $\pi$-$\nu$ correlations. For the 2p-2h $0^{+}$ intruder states, this is expressed \cite{Hey_NPA_87} as
\begin{equation}
E_{intr}(0^{+})=2(\varepsilon_{p}-\varepsilon_{h})-\Delta E_{pairing}+\Delta E_{\pi \nu},
\label{eq:Eintr}
\end{equation}
where $E_{intr}(0^{+})$ is the excitation energy of the $0^{+}$ intruder state, $\varepsilon_{p}-\varepsilon_{h}$ the single-particle shell-gap energy with the respective subscripts $p$ and $h$ denoting particles and holes, $\Delta E_{pairing}$ the nucleon pairing energy, and $\Delta E_{\pi \nu}$ the $\pi$-$\nu$ residual-interaction energy.

The shell-gap and pairing energies for $^{68}$Ni and $^{90}$Zr are deduced from measured one- and two-nucleon separation energies \cite{Aud_NPA_03,Rah_EPJ_07} (see Ref.~\cite{Hey_NPA_87} for details). Starting from the experimental 2p-2h $0^{+}$ excitation energies, the respective $\pi$-$\nu$ residual energies can be extracted, using equation \ref{eq:Eintr}. These values are listed in Table~\ref{tbl:Intruders}. It is important to note that the $E_{intr}(0^{+})$ values in the table are subject to mixing, and no transfer data are known for $^{68}$Ni. The excitation energy of the $^{90}$Zr $\nu$(2p-2h) configuration, e.g., is taken as the average of the $4126$- and $5441$-keV levels, which are strongly populated in the $^{92}$Zr($p$,$t$) reaction \cite{Bal_PRC_71}.

\begin{table}[hb]
\begin{center}
\caption{$^{68}$Ni and $^{90}$Zr are compared starting from their $E_{intr}(0^{+})$ values arising from 2p-2h excitations across the indicated neutron and proton gaps. Also the corresponding $\varepsilon_{p}-\varepsilon_{h}$, $\Delta E_{pairing}$, and $\Delta E_{\pi \nu}$ values are compared.\label{tbl:Intruders}}
\begin{footnotesize}
\begin{ruledtabular}
\begin{tabular}{cccccc}
 Isotope   & Gap  & $E_{intr}(0^{+})$ & $\varepsilon_{p}-\varepsilon_{h}$ & $\Delta E_{pairing}$ & $\Delta E_{\pi \nu}$  \\
 &      & (keV) & (keV) & (keV) & (keV)  \\
\hline
 $^{68}$Ni & $Z=28$ & 2202\footnote{Estimate from summing $\pi$(2p-1h) and $\pi$(1p-2h) excitation energies.} & 5270(320) & 4500(700) & -3838(1000)  \\
\hline
 $^{68}$Ni & $N=40$ & 1770 & 3050(100) & 4705(14) & 380(200)     \\
\hline
 $^{90}$Zr & $Z=40$ & 1761 & 2670(90) & 3593(8) & 20(190)  \\
\hline
 $^{90}$Zr & $N=50$ & 4126 & 4445(8) & 4093(12) & -670(20)  \\
           &        & 5441 & 4445(8) & 4093(12) & 640(20)  \\
           &    & av. 4784 & 4445(8) & 4093(12) & -10(20)  \\
\end{tabular}
\end{ruledtabular}
\end{footnotesize}
\end{center}
\end{table}

\begin{figure}
\centering
\includegraphics[width=0.65\linewidth]{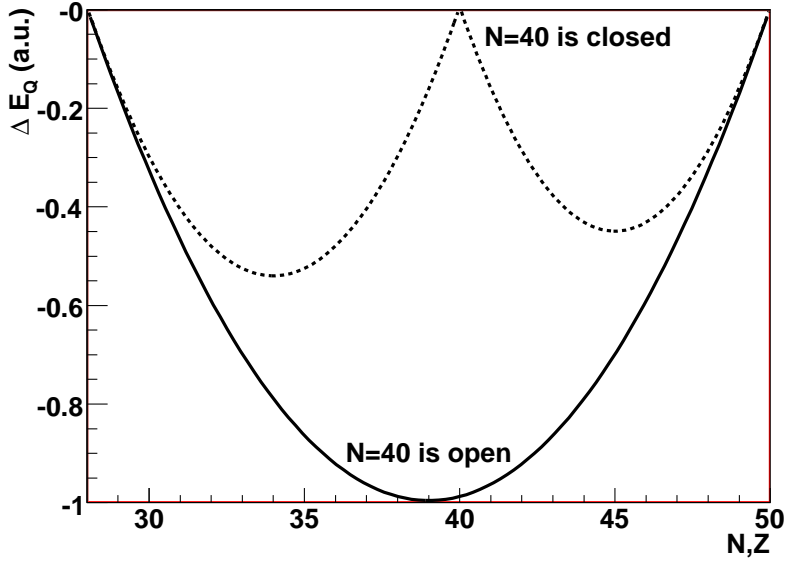}
 \caption{\small{Schematic representation of the quadrupole $\pi$-$\nu$ energy $\Delta E_{Q}$ \cite{Hey_NPA_87} as a function of neutron (proton) number between the closed shells at $N(Z)=28$ and $N(Z)=50$ assuming two extreme cases: $N(Z)=40$ is a closed (dashed lines) and $N(Z)=40$ is an open (full line) shell configuration.}}
 \label{fig:ZN40_EQ}
\end{figure}

The extracted $\pi$-$\nu$ residual-interaction energy mainly results from quadrupole correlations. Fig.~\ref{fig:ZN40_EQ} shows a schematic representation of the quadrupole $\pi$-$\nu$ energy $\Delta E_{Q}$ \cite{Hey_NPA_87} as a function of neutron (proton) number between the closed shells at $N(Z)=28$ and $N(Z)=50$ assuming two extreme cases: $Z(N)=40$ is a closed (dashed lines) and open (full line) shell configuration. In the latter case, the contribution of quadrupole correlations is strongest around $N=39$, and intruder states are expected lowest in excitation energy. On the other hand, if $Z(N)=40$ represents a shell closure, the contribution of quadrupole correlations becomes negligible around $Z(N)=39$, and pairing-excitation states at high excitation energy might be observed.

\subsection{Discussion}

The $\nu$(2p-2h) and $\pi$(2p-2h) $0^{+}$ states in $^{68}$Ni reside at respective excitation energies of $1770$ keV and $\sim 2.2$ MeV, which are rather similar, even though the $Z=28$ shell gap is about $2.2$ MeV larger than the $N=40$ subshell gap. For both gaps, a large gain in pairing energy ($4500$ and $4705$ keV, respectively) exists. For $N=40$, it fully explains the low $\nu$(2p-2h) excitation energy. The low excitation energy of the $\pi$(2p-2h) state, on the other hand, requires a strong gain in binding energy from the $\pi$-$\nu$ residual interactions ($-3.8(10)$ MeV). This means that many valence neutrons must be available, i.e., $N=40$ tends to behave rather as an open shell configuration, as given by the full line in Fig.~\ref{fig:ZN40_EQ}.

As noticed already, the $\pi$(2p-2h) $0^{+}$ state in $^{90}$Zr appears at a remarkably similar excitation energy to the $\nu$(2p-2h) state in $^{68}$Ni. Table~\ref{tbl:Intruders} reveals, however, that the larger shell-gap energies at $N=40$ compared to $Z=40$ is mainly compensated by a stronger gain in pairing energy. So, although both excitation energies are almost identical, the situations at $N=40$ and $Z=40$ are different. Like the $\pi$(2p-2h) state in $^{90}$Zr, the low excitation energy of the $\nu$(2p-2h) $0^{+}$ state in $^{68}$Ni is explained by the gain in pairing energy, which is consistent with a good $Z=28$ shell closure.

The $\nu$(2p-2h) configuration in $^{90}$Zr is centered at a significantly higher excitation energy ($4784$ keV) than the $\pi$(2p-2h) state in $^{68}$Ni ($2.2$ MeV), despite a $0.8$-MeV smaller $N=50$ shell gap and similar pairing energy. This implies a much weaker $\pi$-$\nu$ residual interactions in the $\nu$(2p-2h) states of $^{90}$Zr: the average excitation energy of $4784$~keV is consistent with essentially no $\pi$-$\nu$ residual interaction. In contrast to $N=40$ in the nickel isotopes, $Z=40$ behaves as a closed shell configuration, as depicted in Fig.~\ref{fig:ZN40_EQ} by the dashed lines.

It can be seen from Table~\ref{tbl:Intruders} that the open and closed character as observed in the $0^{+}$ properties of the $N=40$ and $Z=40$ subshell is caused by a stronger pair scattering of neutrons across $N=40$ than protons across $Z=40$: at $N=40$, the pairing energy is about $1.65$ MeV larger than the shell gap, while at $Z=40$, this amounts only to about $0.9$ MeV. Moreover, the difference in pairing energies compensates the difference in unperturbed shell-gap energies giving rise to almost identical excitation energies of the $0^{+}_{2}$ states in $^{68}$Ni and $^{90}$Zr.

In $^{71,73}$Cu, the $\pi$(2p-1h) $7/2^{-}$ levels are identified at $981$ and $1010$ keV, respectively, based on the particle-core model \cite{Oro_NPA_00} and the small B(E2) transition strength \cite{Ste_PRL_08}. This is $\sim 700$ keV lower in excitation energy with respect to the $\pi$(2p-1h) state in $^{69}$Cu. Extrapolating this trend to the nickel and cobalt isotopes, means that the intruder configuration might reside at even lower excitation energies in $^{70,72}$Ni and become even the ground state in $^{69,71}$Co.

\section{Conclusion}

The $^{68}$Ni and $^{90}$Zr low-energy structures have been compared in the framework of 2p-2h configurations across the $Z=28$, $N=40$ and $Z=40$, $N=50$ (sub)shell gaps. The discussion was triggered by recent experimental data obtained in $^{69}$Cu \cite{Ste_PRL_08} and $^{67}$Co \cite{Pau_PRC_08}. Strong similarities are observed between the two valence counterparts, but also important differences. The $0^{+}_{2}$ states in the respective nuclei feature almost identical excitation energies  and monopole $\rho^{2}$(E$0$) transition strengths. Based on the summing prescription of the $\pi$(1p-2h) and $\pi$(2p-1h) levels in $^{67}$Co and $^{69}$Cu, respectively, the $\pi$(2p-2h) $0^{+}$ state in $^{68}$Ni is estimated at only $2.2$-MeV excitation energy, while the $\nu$(2p-2h) $0^{+}$ state in $^{90}$Zr is centered around $4784$ keV.

In an attempt to understand the origin of this difference in $^{68}$Ni and $^{90}$Zr, the shell-model description of $0^{+}$ intruder states \cite{Hey_NPA_87} has been applied. It shows that the excitation energies of the $0^{+}_{2}$ states in $^{68}$Ni and $^{90}$Zr are similar in spite of the difference in the unperturbed single-particle shell-gap energies, as it is compensated to a large extent by the difference in pairing energies. Moreover, it shows that stronger neutron pair scattering in $^{68}$Ni gives rise to more active valence neutrons, which strongly interact with proton excitations across $Z=28$. As a result, the $\pi$(2p-2h) state in $^{68}$Ni is strongly pushed down by $\pi$-$\nu$ residual interactions by as much as $3.8(10)$ MeV, while the $\nu$(2p-2h) state in $^{90}$Zr is hardly affected by $\pi$-$\nu$ residual interactions. These findings highlight the fact that neutron pair scattering across $N=40$ around $^{68}$Ni is far more important than proton pair scattering across $Z=40$ around $^{90}$Zr.

\begin{acknowledgments}
This work was supported by FWO-Vlaanderen (Belgium), GOA/2004/03 (BOF-K.U.Leuven), the 'Interuniversity Attraction Poles Programme -- Belgian State -- Belgian Science Policy' (BriX network P6/23), the
European Commission within the Sixth Framework Programme
through I3-EURONS (contract no. RII3-CT-2004-
506065).
\end{acknowledgments}

\end{document}